\documentclass[%
 reprint,
superscriptaddress,
 amsmath,amssymb,
 aps,
showkeys,
]{revtex4-2}

\usepackage{graphicx,xcolor,soul,siunitx,hyperref}
\graphicspath{{Figures/}}
\usepackage[version=3]{mhchem}
\usepackage{dcolumn}
\usepackage{bm}
\usepackage{cleveref}
\usepackage{multirow}
            
\usepackage{algorithmic}
\usepackage{algorithm}

\newcommand\hil[1]{%
  \bgroup
  \hskip0pt\color{red!80!black}%
  #1%
  \egroup
}

\newif\ifhighlight
\highlightfalse
\ifhighlight
    
\else
    \renewcommand{\hil}[1]{#1}
\fi

\begin{document}

\preprint{Preprint}

\title{Smart structural health monitoring (SHM) system for on-board localization of defects in pipes using torsional ultrasonic guided waves}

\author{Sheetal Patil}
\affiliation{Department of Electrical Engineering, Indian Institute of Technology Bombay, Mumbai 400076, MH India}

\author{Sauvik Banerjee}
\affiliation{Department of Civil Engineering, Indian Institute of Technology Bombay, Mumbai 400076, MH India}

\author{Siddharth Tallur}
\email{stallur@ee.iitb.ac.in}
\affiliation{Department of Electrical Engineering, Indian Institute of Technology Bombay, Mumbai 400076, MH India}


\date{\today}

\begin{abstract}
Most reported research for monitoring health of pipelines using ultrasonic guided waves (GW) typically utilize bulky piezoelectric transducer rings and laboratory-grade ultrasonic non-destructive testing (NDT) equipment. Consequently, the translation of these approaches from laboratory settings to field-deployable systems for real-time structural health monitoring (SHM) becomes challenging. In this work, we present an innovative algorithm for damage identification and localization in pipes, implemented on a compact FPGA-based smart GW-SHM system. The custom-designed board, featuring a Xilinx Artix-7 FPGA and front-end electronics, is capable of actuating the PZT thickness shear mode transducers, data acquisition and recording from PZT sensors and generating a damage index (DI) map for localizing the damage on the structure. The algorithm is a variation of the common source method adapted for cylindrical geometry. The utility of the algorithm is demonstrated for detection and localization of defects such as notch and mass loading on a steel pipe, through extensive finite element (FE) method simulations. Experimental results obtained using a C-clamp for applying mass loading on the pipe show good agreement with the FE simulations. The localization error values for experimental data analyzed using C code on a processor implemented on the FPGA are consistent with algorithm results generated on a computer running MATLAB code. The system presented in this study is suitable for a wide range of GW-SHM applications, especially in cost-sensitive scenarios that benefit from on-node signal processing over cloud-based solutions.
\end{abstract}

                              
\maketitle

\section{Introduction}
Over the course of their long service life, pipelines may undergo metal deterioration, corrosion, and defects due to mass loading caused by external pressure. These factors collectively present substantial risks to the structural integrity of pipelines, potentially resulting in environmental harm and significant financial losses.
In order to mitigate these risks, a variety of non-destructive testing (NDT) techniques are utilized, each providing distinct advantages and valuable insights into the condition of the pipeline \cite{carvalho2008reliability,ma2021pipeline,coramik2017discontinuity}.
Among these methods, ultrasonic guided wave (GW) testing is a highly effective approach for long-range monitoring and inspection of pipelines. Ultrasonic GWs are highly sensitive to early-stage defects and can propagate over extended distances without significant signal distortion and attenuation
\cite{zang2023ultrasonic,ghavamian2018detection}.
Unlike planar structures such as plates, the curved geometry of pipes leads to highly multimodal GW profile comprising of axisymmetric (longitudinal and torsional) and non-axisymmetric (flexural) modes \cite{wang2021time,lowe2015inspection}. Axisymmetric modes are typically utilized for their sensitivity to axial defects, while non-axisymmetric flexural modes may be used for detecting circumferential defects \cite{zhang2017helical}. 
These modes exhibit significant dispersion, and the out-of-plane displacement of the medium during wave propagation leads to decrease in signal amplitude when surrounded by media, like fluid flowing inside the pipe, or external materials such as soil for buried pipes. Unlike these modes, torsional modes are non-dispersive and less affected by the presence of surrounding media, and also highly sensitive to axial and circumferential defects \cite{kim2015review,niu2019excitation}. 

\begin{figure*}[!tbp]
\centering{\includegraphics[width=1\linewidth]{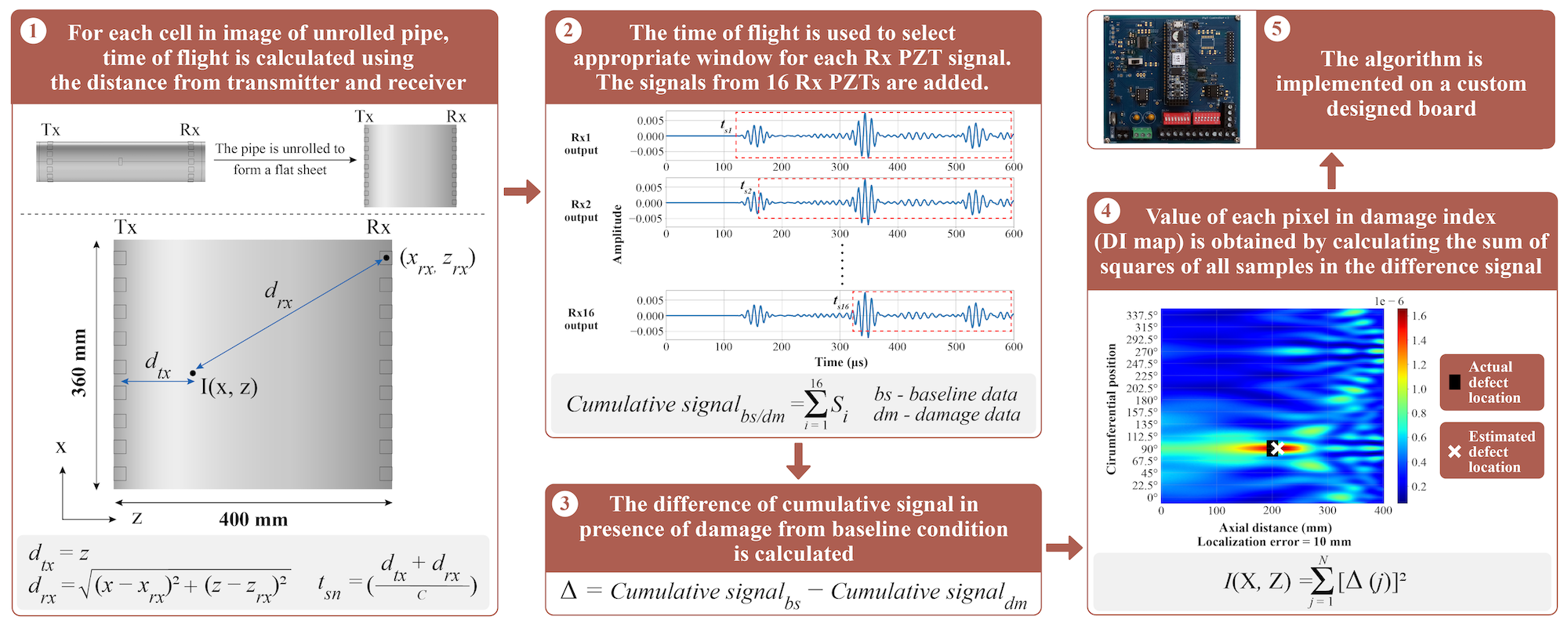}}
\caption{Illustration of the damage localization method implemented on the FPGA Cmod A7.}
\label{fig:damagealgo}
\end{figure*}

Various studies have been reported on utilization of arrays of magnetostrictive and shear mode piezoelectric transducers for actuation of torsional modes for pipe inspection \cite{kim2013circumferential,miao2017excitation,zhou2016guided,heinlein2018reflection}.
These investigations typically involve non-destructive testing on pipe sections using commercially available systems and custom-built laboratory setups. For instance, the Teletest MK3 system has been leveraged for torsional mode based detection of defects in steel pipe by Niu et al. \cite{niu2019excitation}.
Advanced instruments like the Guided Ultrasonic Ltd. Wavemaker G3 
have been employed to study reflections of GWs from from partial and through thickness holes  by Lovstad et al. \cite{lovstad2011reflection}, and for localization of axial defects using common source method (CSM) for synthetic focusing by Fletcher et al. \cite{fletcher2012detection}. 
Using custom-built laboratory setups, other researchers have demonstrated the utility of torsional modes to detect circumferential and longitudinal defects \cite{liu2006circumferential} and observed mode conversion phenomena due to cracks present at different depths in the pipe \cite{yeung2019time}.
However, many of these demonstrations often lead to the development of non-destructive testing and evaluation (NDT\&E) methods, suitable for inspecting smaller pipe sections in the laboratory, or installing the transducer array and instrument on demand in the field. 

The industrial acceptance towards replacement of periodic NDT\&E with permanently deployed GW structural health monitoring (SHM) systems for real-time monitoring and detection of incipient defects in structures in the field has been quite limited.
This is attributed partly to organizational and business case issues, along with significant technical challenges that impede deployment and reliable damage detection and quantification, excellently summarized by Cawley et al. \cite{cawley2021development}. One of the critical requirements for any such adoption involves creating robust transduction and instrumentation systems that can withstand the operational environment while ensuring stability, longevity, and effective connectivity. Furthermore, the data processing system must be integrated into the same system for damage assessment on the edge device in the field, and present easily interpretable data to the operator. Such scalability becomes crucial as automation becomes increasingly essential for managing data from hundreds or thousands of deployments. The laboratory-scale demonstrations reported in many studies discussed earlier are not readily useful for field-deployment.

Field programmable gate arrays (FPGAs) offer design flexibility in GW-SHM, by enabling implementation of reconfigurable hardware and signal processing blocks in the system architecture, without hardware redesign. The high speed clock by most FPGAs also makes them suitable for interfacing with high speed digital to analog converters (DACs) and analog to digital converters (ADCs) for the GW transmitters and receivers, respectively \cite{sawant2022temperature,ramana2023effect,kashyap2023tinyml}. 
Recently, we reported a fully-integrated solution using a Xilinx Artix-7 FPGA for data acquisition and control, and edge-inference of damage \cite{kashyap2024unsupervised}. The system leverages TinyML framework for development of light-weight machine learning models that could be directly deployed on the Artix-7 FPGA. However, the unsupervised learning model described in our previous research could only identify the presence of damage within the structure, without the capability to pinpoint its location.
In this work, we present a novel damage identification and localization algorithm for pipes, based on a variation of CSM, adapted for cylindrical geometry. The algorithm is extensively validated for added mass (mass loading) and notch defects on a steel pipe through finite element (FE) simulations. The structure analyzed in this work comprises of a steel pipe instrumented with rings of $d_{15}$ thickness shear transducers for selective transduction of fundamental torsional i.e., T$(0,1)$ mode. The FE simulation results were also experimentally verified for mass loading, with the algorithm deployed and executed on the FPGA. The methodology of the damage localization algorithm presented in this work is shown in Figure \ref{fig:damagealgo}.
The key contributions of this work are:

\begin{itemize}

    \item \textit{Innovative algorithm for damage identification and localization:} We introduce a novel algorithm based on the common source method, specifically designed for identification and localization of defects in pipes using ultrasonic GWs, validated for added mass and notch defects through FE simulations.

    \item \textit{Algorithm deployment for on-board damage localization on FPGA:} The algorithm is deployed on a custom-designed board featuring a Xilinx Artix-7 FPGA and front-end electronics. The system is capable of end-to-end SHM, including actuation of PZT transducers, data acquisition and recording from PZT sensors, and DI map generation for localization of damage. The damage localization algorithm is implemented in C code on the built-in MicroBlaze $32-$bit processor core on the FPGA.

    \item \textit{Experimental validation:} Through FE simulations and experimental results obtained using a C-clamp for mass loading on a steel pipe, we demonstrate the effectiveness of this system in detecting and localizing defects. The FPGA device consumes only \SI{0.3}{W} power and the on-board computation time for DI map generation using limited resources available on the Artix-7 FPGA is approximately \SI{24}{\minute}. The localization error values from experimental data analyzed on the FPGA processor align well with results generated on a computer running MATLAB code.
    
\end{itemize}

\section{Algorithm for damage localization in pipe}

In ultrasonic GW testing, detection and sizing of defects are typically conducted using phased array techniques \cite{satyanarayan2007simulation, zhou2018reference}. Phased arrays facilitate the precise focusing of energy at specific locations on the structure, through the application of calculated time delays and amplitude adjustments to transducers, and could be used to scan a large area. In contrast, synthetic focusing techniques utilized post-processing of signals after data acquisition, needing only a single acquisition for generating a damage index (DI) map for defect localization, thus reducing hardware cost and scanning time overheads \cite{davies2009application}.
Various synthetically focused imaging algorithms have been conventionally employed for damage detection, including the total focusing method (TFM), synthetic aperture focusing technique (SAFT) and CSM \cite{davies2006review}. In TFM, each transducer element is sequentially actuated, and data is recorded at each of the other elements, leading to long data acquisition and processing times due to multiple transmission-reception cycles for each element. SAFT leverages pulse-echo data from each transducer element in turn. Since SAFT and TFM involve actuation of only one transducer at a time, they are not effective at selective transduction of only torsional modes in pipes, and may suffer from errors stemming from other modes. On the other hand, CSM is more suitable, since it uses a common source for excitation of all transmitter elements, and then records signals on all receiver elements.


In this work, we use a variant of the CSM algorithm adopted for a structure with cylindrical geometry. An overview of the algorithm psuedo-code is shown in Algorithm \ref{alg:algo}.
Since we utilize the fundamental torsional T$(0,1)$ mode, all transmitters are actuated concurrently, and data are simultaneously recorded from each receiver PZT. The pipe is metaphorically unrolled and visualized as a flat sheet, with its length representing the distance between the transmitter (Tx) and receiver (Rx) transducer rings, and its height representing the circumference of the pipe. Subsequently, the algorithm generates an image for this flat sheet, with each pixel value computed as per the steps outlined in the flow chart. The subsequent analysis steps require data to be collected in baseline (i.e., defect-free) condition, against which the data collected in presence of defect are compared. The distance of each pixel from the transmitter ring $(d_{tx})$ and each receiver $(d_{rx})$ is computed. Given that the transmitters are excited simultaneously, only the $z$ distance is considered for distance from the transmitter ring, as shown in step $1$ in Figure \ref{fig:damagealgo}. Time-of-flight $\left( T_{tof}=\frac{d_{tx} + d_{rx}}{C}\right)$ values are computed for the pixel for each Rx PZT, where $C=$ \SI{3130}{m/s} is the group velocity of the T(0,1) mode on a steel pipe considered in this work. The time of flight represents the time required for the signal to travel from the pixel to the receiver PZT. Should the pixel correspond to the location of the defect, the deviation in the signal (relative to baseline condition) needs to be amplified.
This is achieved by considering the GW data in a suitable time window from each receiver for further analysis. The index marking the start of each window $\left(sn=T_{tof} \times f_s\right)$ is computed from the time of flight and sampling rate $(f_s)$ used for data acquisition. Data from such windows for all Rx PZTs are then aggregated to form a single array for each pixel. Baseline subtraction is performed, and the sum of squares of the data points in the result is then assigned to the pixel value of each pixel in the DI map. 

\begin{algorithm}[H]
\caption{Calculate pixel values in damage index (DI) map}
\label{alg:algo}

\textbf{Symbols:} 

$R$: number of rows in DI map image

$C$: number of columns in DI map image \\

$N$: number of Rx PZT transducers attached on the pipe

$m$: indices of Rx PZT transducers attached on the pipe

$sn$: sample number marking the beginning of the window

$T_{tof}$: time of flight for guided wave signal 

$f_s$: sampling rate used for data acuisition\\

$d_{tx}$: distance of pixel from transmitter

$d_{rx}$: distance of pixel from receiver

$C$: group velocity of guided wave signal

\hrulefill

\textbf{Inputs:} 

Data collected from $N$ Rx PZT transducers: \\
$baseline[m]$: data collected in baseline condition

$damage[m]$: data collected in presence of damage

\hrulefill

\textbf{Initialize:} 

$window_{bs} \longleftarrow 0$ // baseline

$window_{dm} \longleftarrow 0$ // damage

\hrulefill

\textbf{for all} $i$ $\in$ $R$

\hspace{3mm}\textbf{for all} $j$ $\in$ $C$

\hspace{6mm}\textbf{for all} $m$ 

\hspace{9mm} $t_{bs} \longleftarrow baseline[m]$

\hspace{9mm} $t_{dm} \longleftarrow damage[m]$ \\


\hspace{9mm} \textbf{Calculate \textit{sn} using \textit{tof}:}

\hspace{9mm} $T_{tof} \longleftarrow (d_{tx} + d_{rx}) / C$

\hspace{9mm} $sn \longleftarrow T_{tof} \times f_s$ \\

\hspace{9mm} \textbf{Select window and add data from all windows:}

\hspace{9mm} $window_{bs} \longleftarrow window_{bs}[m] + t_{bs}[sn : sn + 600]$ 

\hspace{9mm} $window_{dm} \longleftarrow window_{dm}[m] + t_{dm}[sn : sn + 600]$ \\


\hspace{6mm} Difference: $\Delta \longleftarrow window_{bs} - window_{dm}$

\hspace{6mm} $DI[i][j] \longleftarrow \sum_{s=1}^{N} \Delta^2$ // assign pixel value in DI map

\end{algorithm}


\begin{table}[!t]
\caption{Parameters of steel pipe used in FE simulation}
\begin{center}
\small
\begin{tabular}{c c}
\hline
\textbf{Parameter} & \multicolumn{1}{c}{\textbf{Value}} \\
\hline
Outer diameter & \SI{114.6}{mm} \\
Thickness & \SI{4}{mm} \\
Length & \SI{1000}{mm} \\
Density & \SI{7850}{kg/m^3} \\
Young's modulus & \SI{200}{GPa} \\
Poisson's ratio & $0.3$ \\
\hline

\end{tabular}
\label{tab:tab_pipe}
\end{center}
\end{table}

\begin{figure}[!t]
\centering{\includegraphics[width=\linewidth]{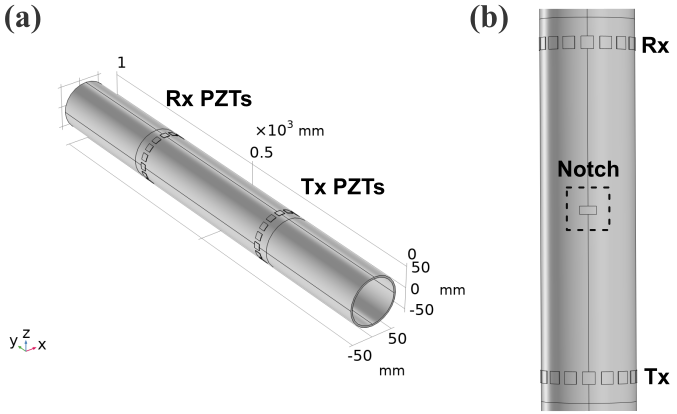}}
\caption{(a) Pipe model used for FE simulations with $16$ transducers for transmission and $16$ transducers for signal reception. (b) Notch defect modeled between the transmitter and receiver PZTs with a circumferential length of \SI{20}{mm}, axial length of \SI{10}{mm} and depth of \SI{2}{mm}.}
\label{fig:sim_pipe}
\end{figure}

\section{Dataset generation for damage localization studies}

\subsection{Finite element simulations}

FE simulations (COMSOL Multiphysics $5.5$) were carried out on a pipe with geometric and material properties as shown in Table \ref{tab:tab_pipe}.
The structure is equipped with shear mode piezoelectric PZT transducers with dimensions \SI{15}{mm} x \SI{15}{mm} x \SI{1}{mm} as shown in Figure \ref{fig:sim_pipe}(a). The material properties for these transducers were the same as the transducers used in experiments (APC International Ltd., 850 WFB), with coordinates specified such that shear waves are generated along the axial direction in the pipe. \textit{Structural Mechanics} module in COMSOL Multiphysics $5.5$ was utilized for modeling the pipe, while the coupling between the pipe and the PZTs was specified through the \textit{Piezoelectricity Multiphysics} interface. 
\textit{User-controlled fine mesh} setting was applied to the pipe geometry, ensuring that the mesh resolution is optimized for each element for wave propagation. \textit{Finer} mesh resolution setting in COMSOL Multiphysics $5.5$ was employed for the transducers, and \textit{extra fine} mesh setting was applied to the portion of the pipe between the transmitter and receiver PZT rings. This meshing strategy aims to balance computational efficiency with modeling precision. The model comprised of \SI{552813}{} degrees of freedom.
The simulation adopted a time-dependent study with a total duration of \SI{600}{\micro s} and a time step of \SI{0.1}{\micro s}. To induce torsional modes, all transmitter PZTs were concurrently excited using a $5-$cycle Hanning windowed sinusoidal pulse:

\begin{equation}
H(t) = 2.5\left(1-\cos\left(\frac{2\pi f t}{5}\right)\right)\sin(2 \pi f t)
\end{equation}
with an excitation frequency of $f=$\SI{85}{kHz}. This voltage profile was applied on the top electrode faces of all $16$ transmitter PZTs, while the bottom surfaces of these PZT transducers were grounded. Output signals were recorded at $16$ receiver PZTs positioned uniformly in a ring spaced \SI{400}{mm} away from the transmitter PZT ring. The generation of torsional modes using this configuration, and optimization of the transduction frequency were discussed in our previous work \cite{patil2023embedded}.
Damage in the pipe was modeled in the form of a notch as shown in Figure \ref{fig:sim_pipe}(b), with circumferential length of \SI{20}{mm}, axial length of \SI{10}{mm}, and depth of \SI{2}{mm}. The notch was positioned between the transmitter and receiver PZTs, and data were recorded at all $16$ receiver PZTs for various axial and circumferential positions of the notch.
The experimental setup described in next sub-section made use of $8$ PZTs for transmitter as well as receiver rings, with mass loading (added mass) applied to the pipe to simulate defect. These conditions were also replicated in FE simulations for mass loading. \SI{10}{kg} mass loading on the pipe was modeled on a circular cross-sectional area of \SI{6}{mm} diameter, for various axial positions between the transmitter and receiver rings. The angular position of the mass loading was maintained at \ang{90}, as the choice of the angular position where the pipe is unrolled to generate the DI map is arbitrary. The simulations for added mass are carried out at \SI{75}{kHz} excitation frequency, as cleaner waveforms were obtained in the experiment for this frequency as compared to \SI{85}{kHz}. 

\begin{figure}[!t]
\centering{\includegraphics[width=0.9\linewidth]{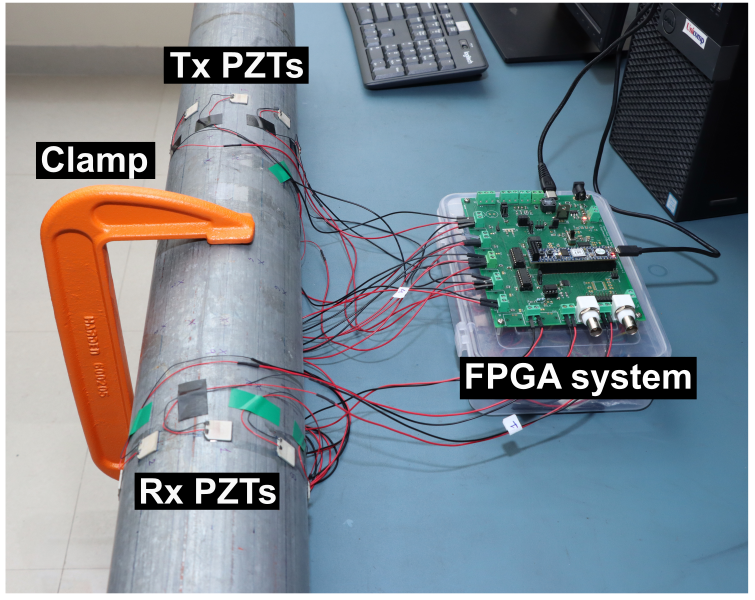}}
\caption{Photograph of experimental setup showing steel pipe fitted with shear mode PZT transducers, C-clamp for applying mass loading on the pipe and FPGA-based smart SHM system for data acquisition and processing.}
\label{fig:exp_setup}
\end{figure}

\subsection{Experimental method}
The experimental setup consists of a steel pipe with dimensions similar to those used in simulations, fitted with shear mode $d_{15}$ PZT transducers with dimensions \qtyproduct{15x15x1}{mm} (APC International Ltd., 850 WFB). Eight PZTs were used as transmitters, and an additional eight were used as receivers, positioned at a distance of \SI{400}{mm} from the transmitter ring. The actuation and data acquisition processes are managed through an FPGA-based embedded system. A photograph of the experimental setup is shown in Figure \ref{fig:exp_setup}. The system uses Xilinx Artix®-7 FPGA (Digilent Cmod A7 module). The FPGA is used for generating a $5$-cycle Hanning windowed sinusoidal pulse at \SI{75}{kHz} frequency. The digital signal is converted to analog using a Texas Instruments DAC7821 $12$-bit parallel input digital-to-analog converter (DAC). Due to the unbuffered current output nature of the DAC7821, a transimpedance amplifier assembled using the Texas Instruments TL072 operational amplifier IC is connected to the DAC output, with suitable gain for achieving a peak-to-peak amplitude of \SI{10}{V} for the excitation pulse.
The signals at the receiver PZTs are amplified using Texas Instruments INA128 instrumentation amplifier and then digitized through the Maxim Integrated MAX1426 analog-to-digital converter (ADC), which has $10-$bit resolution and supports sampling rate as high as \SI{10}{Msps}. The FPGA-based system supports $8$ PZT channels and can store data for up to \SI{800}{\micro s} in its built-in memory.
In the experimental setup, the transmitter PZTs are simultaneously excited with a $5$-cycle Hanning windowed sinusoidal pulse at \SI{75}{kHz}, and data from all eight receiver PZTs are recorded in the FPGA memory. To reduce noise due to electromagnetic interference (EMI) from ambient noise sources, the system performs averaging for data collected from each PZT, wherein $10$ traces are averaged and recorded in each measurement. The data saved in the FPGA memory is then analyzed, and the results (DI map pixel values) are transmitted to a host computer for visualization via universal asynchronous receiver / transmitter (UART) protocol. The FPGA system architecture is described in detail in the next section. C-clamp is used for applying mass loading on the pipe specimen.

\section{FPGA system architecture}

\begin{figure}[!t]
\centering{\includegraphics[width=\linewidth]{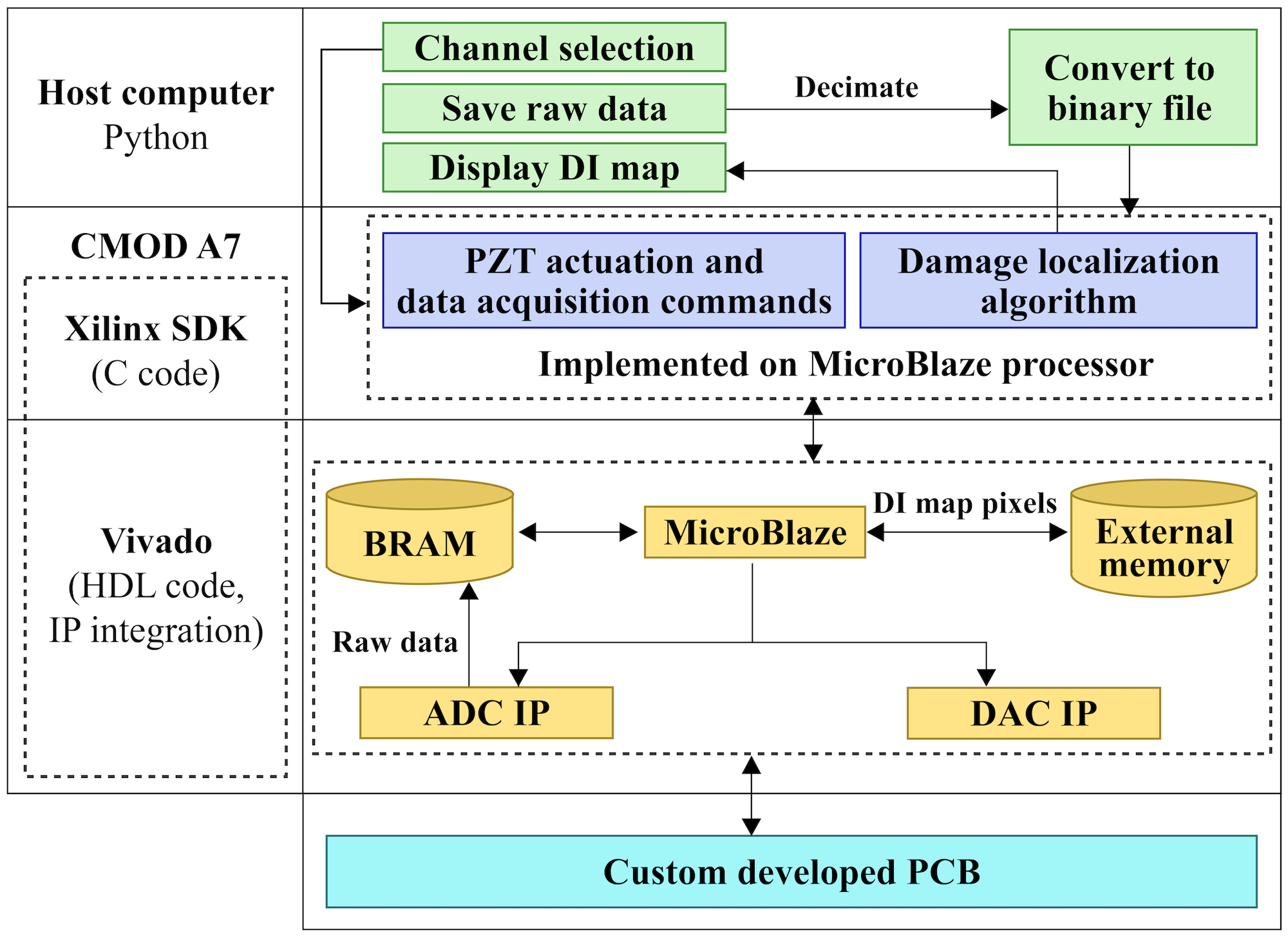}}
\caption{Overview of FPGA system architecture, higlighting various software tools and development environments used for implementing the design.}
\label{fig:system_flowchart}
\end{figure}

An overview of the FPGA system architecture is shown in Figure \ref{fig:system_flowchart} and the resource utilization on the FPGA is presented in Table \ref{tab:resources}. The versatile Cmod A7 module with Xilinx Artix-7 FPGA contains \SI{512}{KB} static random access memory (SRAM), \SI{225}{KB} block random access memory (BRAM), \SI{4}{MB} quad-SPI flash memory, a $48-$pin DIP connector for simplified connections to other electronic components, USB-JTAG programming circuit for device configuration and debugging, and  USB-UART bridge for communication with host computer.
The hardware implemented in the FPGA is configured using Vivado Design Suite, a tool utilized for the synthesis and analysis of digital systems. 
The MicroBlaze ($32-$bit RISC soft processor core optimized for embedded applications on Xilinx devices) is the central element of our architecture for controlling the execution flow of the hardware.
We have used Vivado HLS tool along with Vivado Design Suite for hardware modeling and implementation, and Xilinx software development kit (SDK) to design an embedded application for controlling and interfacing the implemented hardware with external peripherals. Xilinx SDK is an integrated development environment within the Xilinx Vivado Design Suite, providing tools for cross-platform software development, debugging, and code generation for embedded systems running on Xilinx FPGAs and SoCs. It involves programming the processor in C language to implement commands to enable and route signals to various built-in and custom designed intellectual property (IP) blocks for initiating signal transduction, and executing algorithms for damage localization. The design includes custom IP blocks for interfacing with the ADC and DAC ICs.

\begin{table}[!t]
\caption{Resource utilization on FPGA module}
\begin{center}
\small
\begin{tabular}{c c c c c}
\hline
\textbf{Microblaze}& \centering \textbf{Flipflops} & \centering \textbf{LUTs} &\centering \textbf{BRAM}&\multicolumn{1}{c}{\textbf{DSP slices}} \\
\cline{1-5} 
Available & \SI{41600}{} & \SI{20800}{} & $50$ & $90$ \\
Utilized & $4276$ & $5314$ & $35.5$ & $2$ \\
\hline
\end{tabular}
\label{tab:resources}
\end{center}
\end{table}

\begin{figure*}[!t]
\centering{\includegraphics[width=0.7\linewidth]{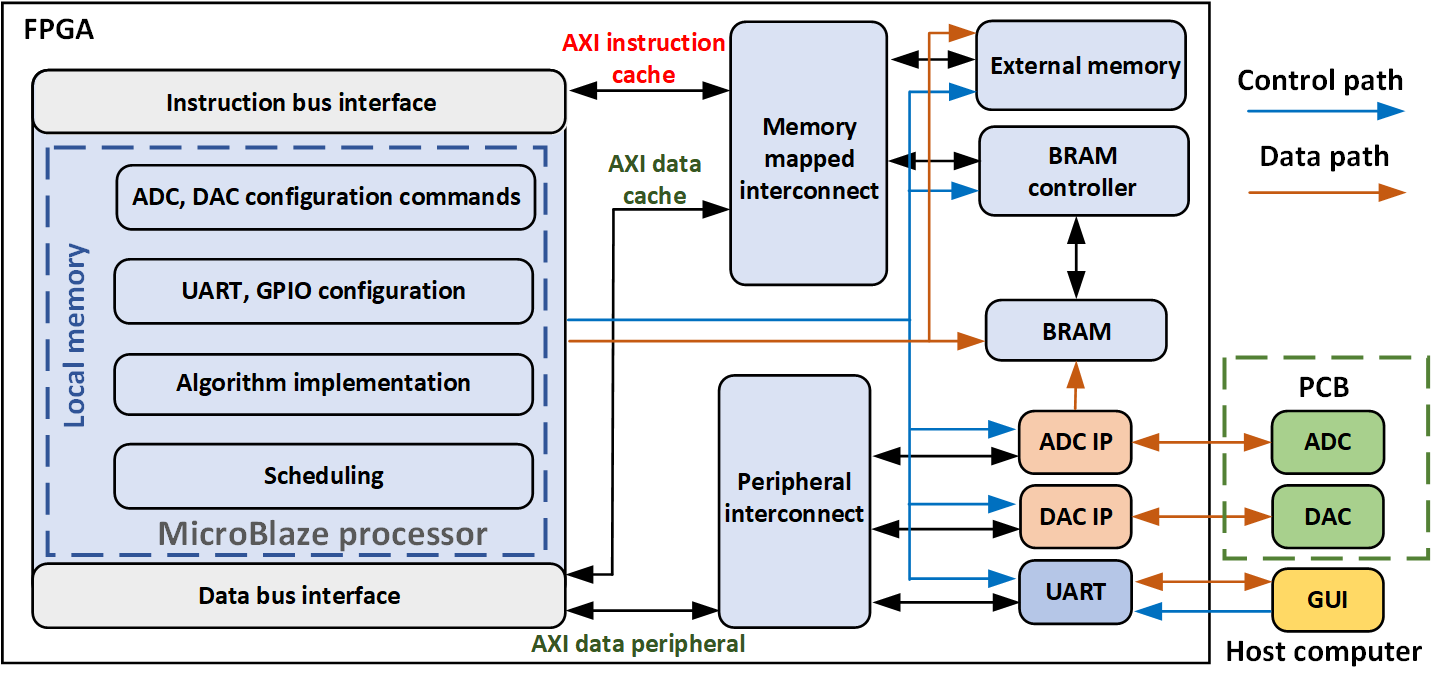}}
\caption{FPGA architecture highlighting interconnects, control and data paths.}
\label{fig:dataflow}
\end{figure*}

A Python-based graphical user interface (GUI) was implemented and run on a host computer to configure and acquire data from the FPGA. Utilizing the FTDI FT2232HQ USB-UART bridge available on the Cmod A7 module, communication between PC application and the board is facilitated through standard Windows COM port commands. The GUI provides functionality for channel selection and UART-based data acquisition, involving the capture of $6000$ samples per channel. This amounts to \SI{48000}{} samples for $8$ PZT Rx transducers in each of baseline and damage conditions, resulting in a total of \SI{96000}{} samples. To handle this substantial data volume, The signals are decimated by a factor of $10X$, reducing the dataset to $9600$ samples that can fit comfortably in the memory available on the board. For separately running the damage assessment algorithm only on the FPGA, data recorded offline can be converted to binary format and directly uploaded to the FPGA memory using the Xilinx software command line tool (XSCT).
The damage localization algorithm is implemented using C code, and the computation is performed using the MicroBlaze processor core. The pixel values for the DI map thus computed are stored in the external memory of the FPGA for subsequent retrieval to host computer for visualization. The algorithm computation time was experimentally determined to be approximately \SI{24}{\minute} on the FPGA, and could be sped up by upgrading to an FPGA with more resources, or decimating the data even further. The power consumption of the FPGA device was experimentally determined to be approximately \SI{0.31}{W}. 

\subsection{MicroBlaze processor program architecture}

The MicroBlaze processor code architecture and its connections with memory blocks and peripherals are shown in Figure \ref{fig:dataflow}. At onset of system operation, essential boot code residing in a dedicated local memory is accessed with a single-cycle latency for swift instruction and data access. Operating the processor exclusively from local memory eliminates the need for caches, reducing the resource overhead of the MicroBlaze core. This dedicated local memory is inaccessible to other peripheral components.
To facilitate memory access for both the MicroBlaze processor and custom IPs, a shared memory location in BRAM is integrated into the design. This BRAM connects to the MicroBlaze through cached interfaces, including the Data cache and Instruction cache, connected through a memory-mapped interconnect. The interconnect introduces access latency, necessitating the use of caches for efficient operation. Additionally, given the substantial volume of output data generated by the algorithm, an external memory interface to on-board memory is incorporated into the system architecture.
Note that the entire processor code can also be implemented in the external memory for algorithms with higher memory requirement, but this increases the instruction access latency which may in turn increase the computation time.
The MicroBlaze serves as the master controller for all peripherals, orchestrating control signals for various IPs, including initiation and termination commands for both the DAC and ADC modules. Specifically, it triggers the DAC to produce the excitation signal to the transmitter PZTs by issuing start commands, and manages the data acquisition process by initiating ADC operations and controlling data storage. Additionally, the MicroBlaze handles UART initialization and configuration, enabling communication with the host computer running the Python-based GUI program. 

\begin{figure*}[!t]
\centering{\includegraphics[width=1\linewidth]{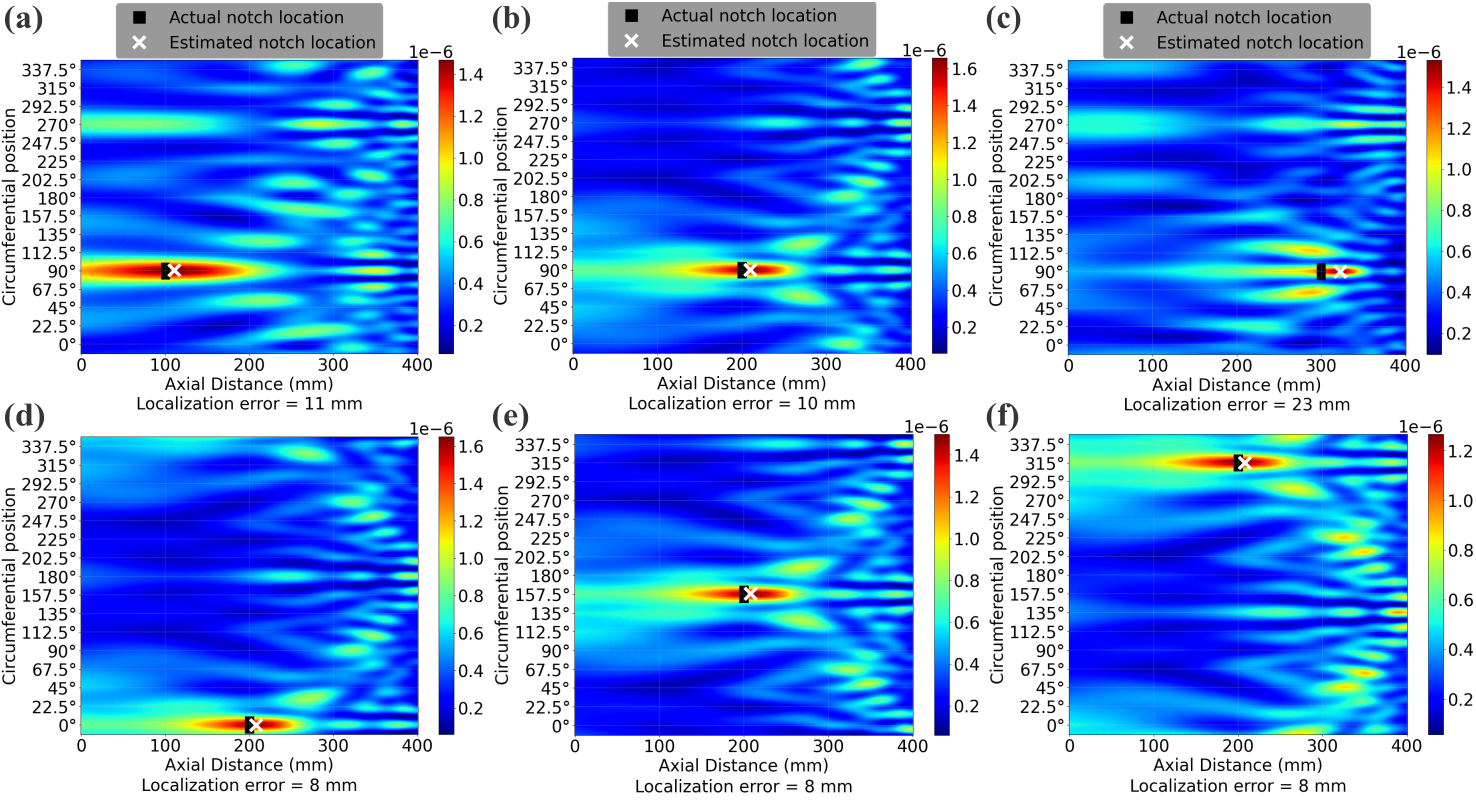}}
\caption{DI maps for notch defect placed at various angular and axial locations on the pipe in FE simulations. The actual notch location and the value estimated from the algorithm are shown in each DI map. For angular position of \ang{90}, the location of notch is varied in axial direction for separation from transmitter ring equal to (a) \SI{100}{mm}, (b) \SI{200}{mm}, and (c) \SI{300}{mm}. Next, for axial position of \SI{200}{mm} from transmitter ring, angular position of the notch is set equal to (d) \ang{0}, (e) \ang{157.5}, and (f) \ang{315}. }
\label{fig:sim_result}
\end{figure*}

\begin{figure*}[!t]
\centering{\includegraphics[width=1\linewidth]{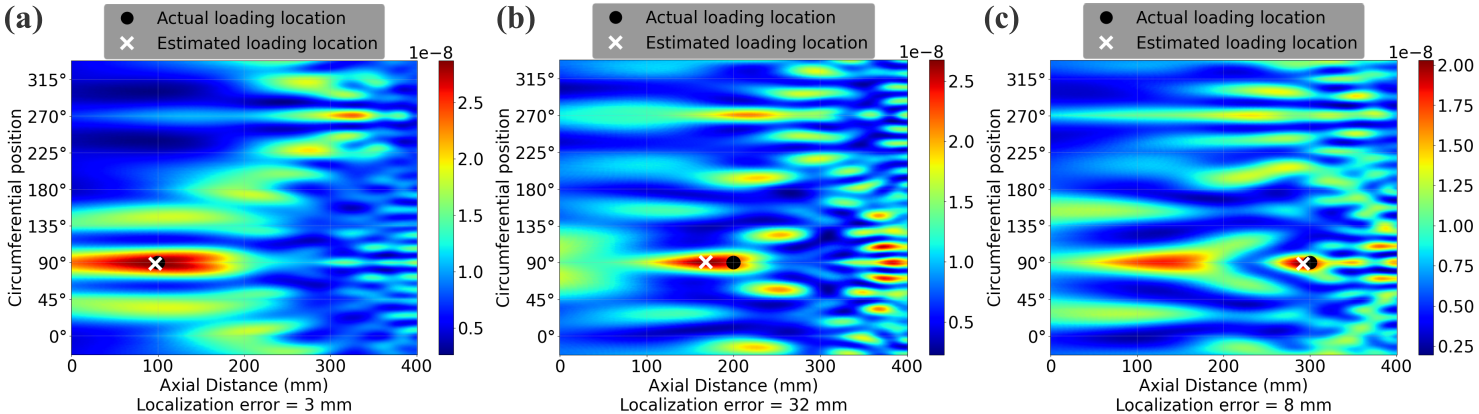}}
\caption{DI maps for added mass placed at various axial locations on the pipe in FE simulations. The actual added mass location and the value obtained from the algorithm are shown in each DI map. For angular position of \ang{90}, the location of added mass is varied in axial direction for separation from transmitter ring equal to (a) \SI{100}{mm}, (b) \SI{200}{mm}, and (c) \SI{300}{mm}.}
\label{fig:sim_resultam}
\end{figure*}

\subsection{Peripherals}

Custom-designed IPs, developed utilizing VHDL code, facilitate seamless communication with the on-board ADC and DAC ICs. The DAC IP is responsible for generating the excitation signal for driving the transmitter PZTs. This functionality is achieved through the generation of digital codes stored in look-up tables (LUTs), that are transmitted to the on-board DAC and precisely reconstruct the desired excitation signal at the DAC output. 
The ADC IP is designed to manage the control signals for the on-board ADC, and efficiently storing the data acquired from the receiver PZTs without data loss. On-board electronic switches are used to connect appropriate receiver channel to the ADC for sequential data acquisition. The data thus acquired are stored in BRAM for subsequent processing.
UART IP configured with a baud rate of \SI{230400}{} is used for transmitting acquired data, DI map pixel values and system information between the FPGA and the host computer. 

\begin{figure*}[!t]
\centering{\includegraphics[width=1\linewidth]{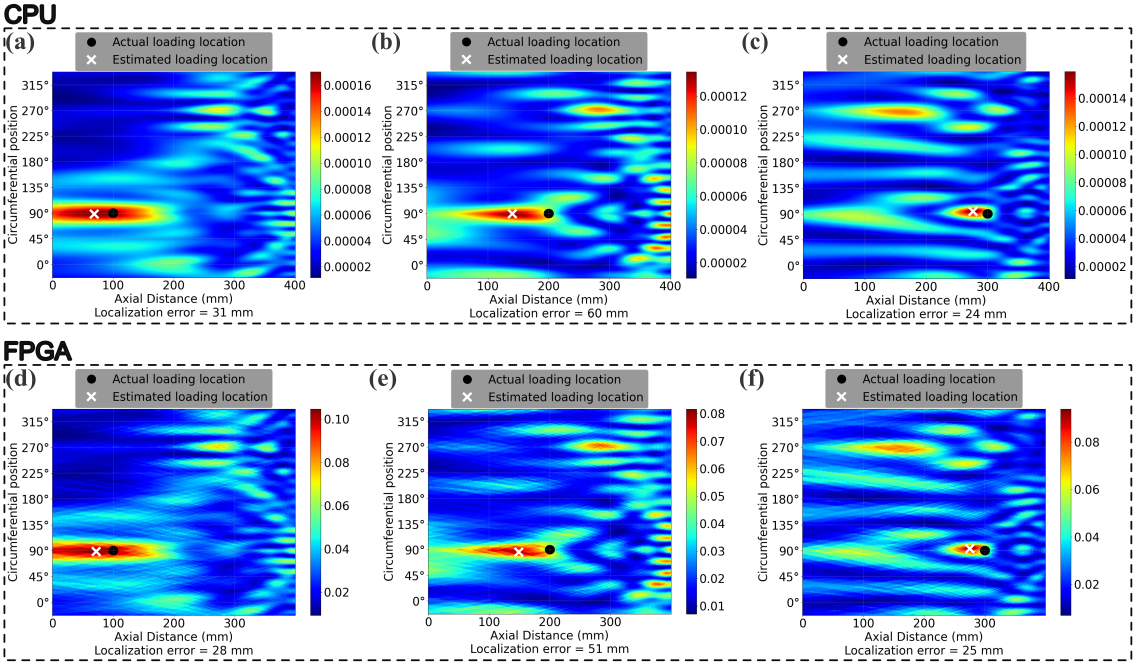}}
\caption{DI maps for mass loading applied at various axial locations on the pipe using C-clamp in experimental setup.  The actual loading location and the one obtained from the algorithm are shown in each DI map. For angular position of \ang{90}, the location of added mass is varied in axial direction for separation from transmitter ring equal to (a, d) \SI{100}{mm}, (b, e) \SI{200}{mm}, and (c, f) \SI{300}{mm}. Panels (a)-(c) show results obtained by running the damage localization algorithm on a CPU computer, while panels (d)-(f) show results obtained from implementing the algorithm on FPGA.}
\label{fig:exp_result}
\end{figure*}

\section{Results and discussion}

\subsection{FE simulation results}

The DI map results generated using Algorithm \ref{alg:algo} for various simulations performed for notch defect and added mass are shown in Figures \ref{fig:sim_result} and \ref{fig:sim_resultam}, respectively. The vertical axis in each DI map represents the circumferential (i.e., angular) positions along the pipe, while the horizontal axis denotes the axial distance between the transmitter and receiver PZT rings. Figures \ref{fig:sim_result}(a)-(c) show results obtained for notch defect positioned at \ang{90}, whose axial position was varied such that the separation from transmitter ring is \SI{100}{mm}, \SI{200}{mm} and \SI{300}{mm}, respectively. The localization error was computed by measuring the distance between the actual notch location as specified in the simulation, and the estimated notch location determined by the algorithm.
Next, with the axial position held constant for \SI{200}{mm} separation from transmitter ring, the angular position of the notch was varied to \ang{0}, \ang{157.5} and \ang{315}, resulting in DI maps shown in Figures \ref{fig:sim_result}(d)-(f), respectively. 
Despite the small size of the notch and its partial through-hole nature, it causes measurably large changes in the receiver signals. The algorithm successfully localizes the notch position, even when it is near the transmitter or receiver, albeit with some error. 
A maximum localization error of \SI{23}{mm} was observed for a notch located closer to the receiver PZTs at angular position of \ang{90}, as seen in Figure \ref{fig:sim_result}(c). The magnitude of this localization error is small relative to the \SI{400}{mm} separation between the transmitter and receiver PZT rings.
DI map results for added mass positioned at three different axial locations on the pipe (\SI{100}{mm}, \SI{200}{mm}, and \SI{300}{mm} separation from transmitter ring) are shown in Figures \ref{fig:sim_resultam}(a)-(c). 
The algorithm effectively localizes the added mass with low error. 

\begin{figure*}[!t]
\centering{\includegraphics[width=1\linewidth]{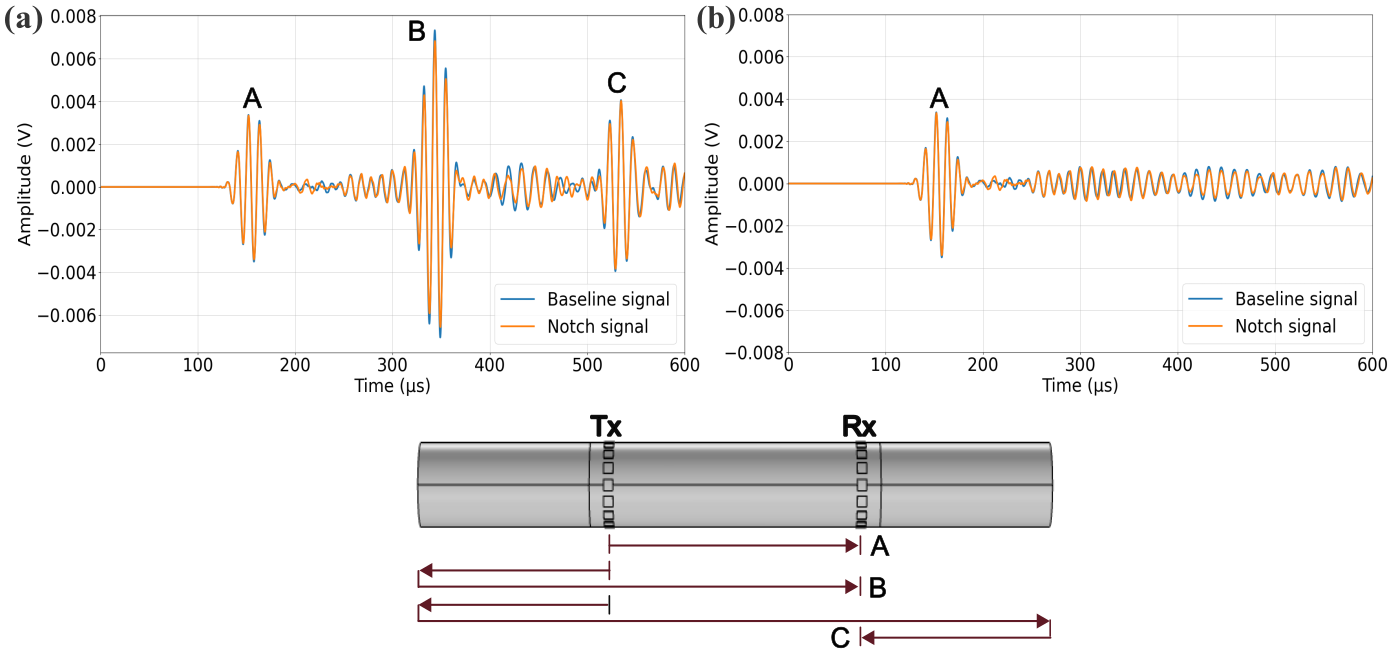}}
\caption{Representative waveforms of GW data recorded at one of the receivers for FE simulations in baseline condition and with notch defect: (a) without, and (b) with low-reflecting boundary condition applied at the axial edges of the pipe. The wavepacket denoted by A is the fundamental T$(0,1)$ mode as received at Rx, while components B and C are reflections of this mode from the edge of the pipe, as shown in the illustration at the bottom of the figure.}
\label{fig:sim_timeseries_notch}
\end{figure*}

\begin{figure*}[!t]
\centering{\includegraphics[width=1\linewidth]{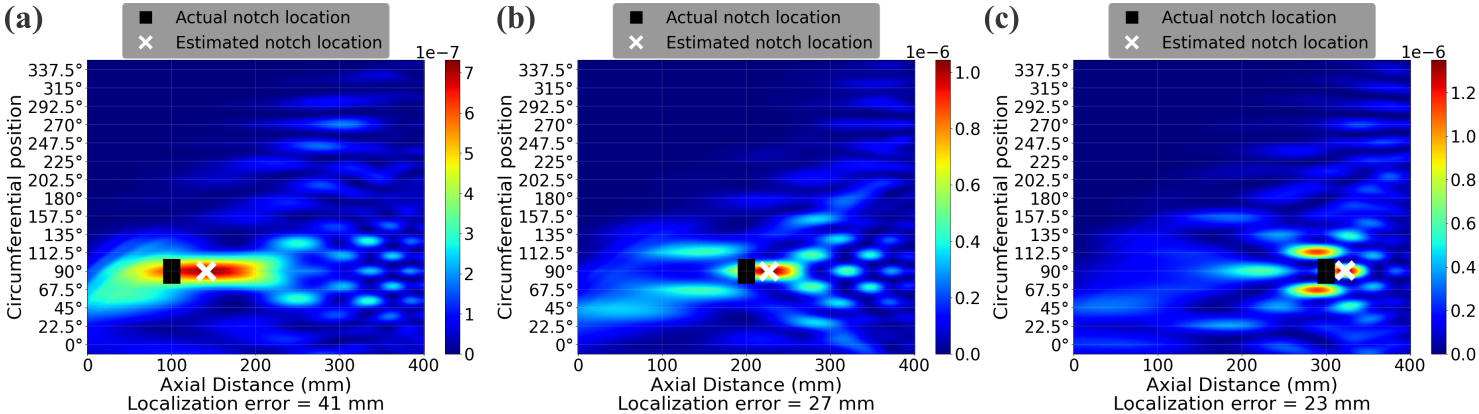}}
\caption{DI maps for notch placed at various axial locations in FE simulations with low-reflecting boundary condition applied at edges of the pipe. The actual location of the notch specified in the simulation and the value obtained from the algorithm are shown in each DI map. For angular position of \ang{90}, the location of notch is varied in axial direction for separation from transmitter ring equal to (a) \SI{100}{mm}, (b) \SI{200}{mm}, and (c) \SI{300}{mm}.}
\label{fig:simnotch_lrf}
\end{figure*}

\subsection{Experimental results}

DI maps for mass loading applied at various axial locations on the pipe using C-clamp in experimental setup are shown in Figure \ref{fig:exp_result}.
For angular position of \ang{90}, the location of mass loading is varied in axial direction. Panels (a)-(c) show results obtained by running the damage localization algorithm on a CPU computer using Python, for axial location of the mass loading for separation from transmitter ring equal to (a) \SI{100}{mm}, (b) \SI{200}{mm}, and (c) \SI{300}{mm}. Since the experimental setup only uses $8$ PZT transducers in the transmitter and receiver rings, which do not result in complete suppression of other longitudinal and flexural modes, the localization error is higher for experimental results as compared to FE simulations. However, the magnitude of localization error is small relative to the size of the pipe, and \SI{400}{mm} separation between the transmitter and receiver PZT rings. The localization error was observed to be higher for mass loading applied mid-way between the transducer rings, which was also noticed for simulation results shown in Figure \ref{fig:sim_resultam}.
The same experimentally recorded dataset was used to assess algorithm performance for implementation on the FPGA. The corresponding DI maps are shown in Figures \ref{fig:exp_result}(d)-(f). Despite performing data decimation to accommodate FPGA resource constraints, patterns seen in the DI maps and localization error remain largely unaltered. 
%

\begin{figure*}[!t]
\centering{\includegraphics[width=1\linewidth]{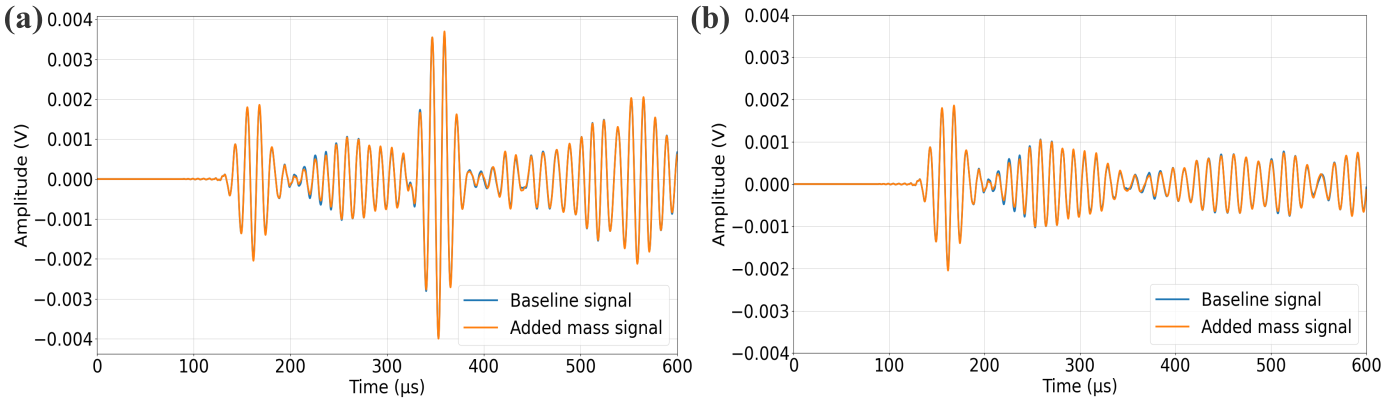}}
\caption{Representative waveforms of GW data recorded at one of the receivers for FE simulations in baseline condition and with added mass: (a) without, and (b) with low-reflecting boundary condition applied at the axial edges of the pipe.}
\label{fig:sim_timeseries_am}
\end{figure*}

\begin{figure*}[!t]
\centering{\includegraphics[width=1\linewidth]{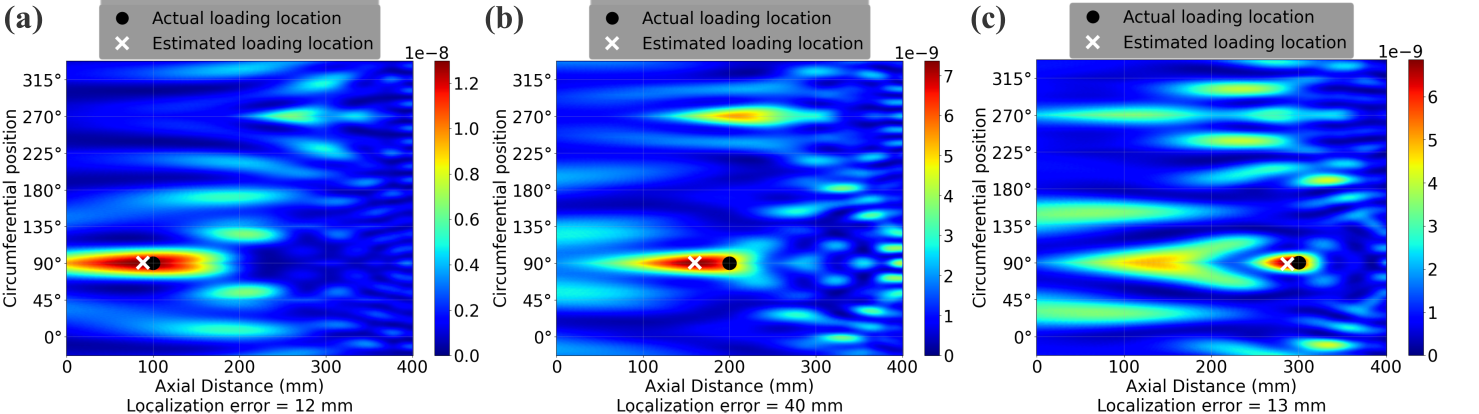}}
\caption{DI maps for added mass placed at various axial locations in FE simulations with low-reflecting boundary condition applied at edges of the pipe. The actual location of the added mass specified in the simulation and the value obtained from the algorithm are shown in each DI map. For angular position of \ang{90}, the location of added mass is varied in axial direction for separation from transmitter ring equal to (a) \SI{100}{mm}, (b) \SI{200}{mm}, and (c) \SI{300}{mm}.}
\label{fig:simam_lrf}
\end{figure*}

\subsection{Discussion}


The FE simulations and experimental setup both result in reflection of the GW mode wavepackets from the boundaries of the pipe. Figure \ref{fig:sim_timeseries_notch}(a) shows representative waveforms of GW data recorded at one of the receivers, showing the fundamental T$(0,1)$ mode and additional components present in the signal in presence of notch defect, due to reflection from the edge of the pipe specimen. The reflected component contributes prominently to the signal difference, and therefore the DI pixel intensity in vicinity of the notch defect in the DI map. Practical specimens will certainly involve reflections from discontinuities such as welded butt-joints, bolted connections, bends etc. Therefore, recording data in baseline condition is imperative for any structure. To check the efficacy of the algorithm in absence of such strong reflected components in the GW signal, we performed additional FE simulations with \textit{Low-Reflecting} boundary condition specified in COMSOL Multiphysics $5.5$ at both edges of the finite-length pipe specimen. Figure \ref{fig:sim_timeseries_notch}(b) shows representative waveform of GW data recorded at one of the receivers, illustrating suppression of the reflected components. Despite this missing component, the algorithm is successfully able to localize the defect, as seen in the DI maps in Figure \ref{fig:simnotch_lrf}. Note that these DI maps are generated using time-series data only corresponding to \SI{200}{\micro s} duration of recording, since there are no reflected components in the signal that need to be accounted for, and because of presence of undesirable longitudinal and flexural modes that are not fully suppressed. For mass loading applied in FE simulation, there is no scattering introduced in the GW signal. Representative waveforms of GW data recorded at one of the receivers with and without low-reflecting boundary condition at the edges of the pipe are shown in Figure \ref{fig:sim_timeseries_am}, and DI maps for FE simulations performed with low-reflecting boundary condition are shown in Figure \ref{fig:simam_lrf}. The introduction of mass loading introduces a very small change in the signal amplitude of the T$(0,1)$ mode. Since only $8$ transducers are employed in transmitter and receiver rings, the longitudinal and flexural modes are not fully suppressed (Figure \ref{fig:sim_timeseries_am}), unlike the case of simulations of notch defect (Figure \ref{fig:sim_timeseries_notch}), which were performed with $16$ transducers in transmitter and receiver rings. Table \ref{tab:tab_ampl} shows signal amplitudes for the T$(0,1)$ mode in baseline condition and three different trials of mass loading applied using C-clamp in the experimental setup. Despite this small change in amplitude, the algorithm is able to successfully localize the position where the mass loading is applied.

\begin{table}[!tbp]
\caption{Signal amplitudes of T$(0,1)$ mode in experimental setup}
\begin{center}
\small
\begin{tabular}{c|c|c}
\hline
\textbf{Test condition} & \textbf{Separation from Tx} & \multicolumn{1}{c}{\textbf{Amplitude}} \\
\hline
Baseline & NA & \SI{82.5}{mV} \\
\hline
\multirow{3}{*}{Mass loading} & \SI{100}{mm} & \SI{81.72}{mV} \\
 & \SI{200}{mm} & \SI{78.2}{mV} \\
 & \SI{300}{mm} & \SI{80.94}{mV} \\

\hline

\end{tabular}
\label{tab:tab_ampl}
\end{center}
\end{table}

The experimental setup could be improved by modifying the clamp to apply a point load instead of an area-distributed load, that could potentially result in better localization error, and enable us to carefully optimize the algorithm further. The density of transducers can also be increased by using smaller transducers, and a power amplifier could be integrated into the system to actuate the transmitters with larger signal power for increasing the range of the system.
By upgrading the FPGA to an alternate device with higher number of resources, data decimation could also be avoided and numerical accuracy of localization could be improved.
It is worth noting that while the algorithm successfully localizes a single damage positioned on the pipe as studied in this work, it is not suitable when applied to specimens containing multiple defects. In future work, we seek to address this limitation by investigating suitable data segmentation, windowing and superposition approaches, in addition to developing machine learning aided algorithms for accurate damage localization in such scenarios. The experimental specimen will be enhanced further to enable testing with pressurized liquid flow in the pipe.



\section{Conclusion and future work}

In summary, this work investigates the application of a novel damage localization algorithm using torsional modes in a metallic pipe, implemented in an embedded GW-SHM system. The system is capable of localizing damage in pipes by itself, wherein signal transduction and processing are all performed on-board using an FPGA device, with tasks controlled using a processor core implemented in the FPGA. The effectiveness of the localization algorithm was validated through FE simulations for notch and added mass applied on the pipe at various axial and angular positions. The algorithm consistently estimated the location of the defect with high across various simulation scenarios. Experimental validation for added mass confirm the effectiveness of the algorithm at localizing defects, albeit with increased localization error due to limitations of the experimental setup.
Furthermore, implementation of the algorithm on the FPGA showed performance comparable to CPU processing, which holds great promise for utility of the smart-SHM system for various GW-SHM applications. 
The Xilinx Artix-7 FPGA used in this study is low-cost and suitable for scalable GW-SHM systems. However, the limited number of resources present in the device as compared to much more advanced devices such as the Xilinx Zynq UltraScale+ MPSoC, results in limitations on computational complexity of algorithms that could be deployed on hardware. These limitations could be overcome by upgrading to the Xilinx Zynq UltraScale+ MPSoC device, and will be explored in future work. This device is also suitable for easily implementing machine learning algorithms by leveraging the built-in deep learning processor unit (DPU) IP, which is not available for the Artix-7 FPGA.
Nevertheless, the system presented in this work could find applications for a variety of practical GW-SHM use cases that require basic signal processing on the sensor node to avoid costs associated with cloud connectivity and storage and post-processing of large volumes of raw data.

\section*{Data availability}
The data cannot be made publicly available upon publication because they are not available in a format that is sufficiently accessible or reusable by other researchers. The data that support the findings of this study are available upon reasonable request from the authors.

\section*{Acknowledgment}
This work was supported through a grant from Centre of Excellence in Oil, Gas and Energy (CoE-OGE), IIT Bombay [grant no. DO/2021-COGE002-003].
The authors acknowledge support from staff and access to facilities at the Wadhwani Electronics Lab (WEL), Department of Electrical Engineering, IIT Bombay for carrying out experiments reported in this work. 


\section*{Declaration of competing interests}
The authors declare that they have no known competing financial interests or personal relationships that could have appeared to influence the work reported in this paper.

\bibliography{main}

\begin{thebibliography}{28}%
\makeatletter
\providecommand \@ifxundefined [1]{%
 \@ifx{#1\undefined}
}%
\providecommand \@ifnum [1]{%
 \ifnum #1\expandafter \@firstoftwo
 \else \expandafter \@secondoftwo
 \fi
}%
\providecommand \@ifx [1]{%
 \ifx #1\expandafter \@firstoftwo
 \else \expandafter \@secondoftwo
 \fi
}%
\providecommand \natexlab [1]{#1}%
\providecommand \enquote  [1]{``#1''}%
\providecommand \bibnamefont  [1]{#1}%
\providecommand \bibfnamefont [1]{#1}%
\providecommand \citenamefont [1]{#1}%
\providecommand \href@noop [0]{\@secondoftwo}%
\providecommand \href [0]{\begingroup \@sanitize@url \@href}%
\providecommand \@href[1]{\@@startlink{#1}\@@href}%
\providecommand \@@href[1]{\endgroup#1\@@endlink}%
\providecommand \@sanitize@url [0]{\catcode `\\12\catcode `\$12\catcode `\&12\catcode `\#12\catcode `\^12\catcode `\_12\catcode `\%12\relax}%
\providecommand \@@startlink[1]{}%
\providecommand \@@endlink[0]{}%
\providecommand \url  [0]{\begingroup\@sanitize@url \@url }%
\providecommand \@url [1]{\endgroup\@href {#1}{\urlprefix }}%
\providecommand \urlprefix  [0]{URL }%
\providecommand \Eprint [0]{\href }%
\providecommand \doibase [0]{https://doi.org/}%
\providecommand \selectlanguage [0]{\@gobble}%
\providecommand \bibinfo  [0]{\@secondoftwo}%
\providecommand \bibfield  [0]{\@secondoftwo}%
\providecommand \translation [1]{[#1]}%
\providecommand \BibitemOpen [0]{}%
\providecommand \bibitemStop [0]{}%
\providecommand \bibitemNoStop [0]{.\EOS\space}%
\providecommand \EOS [0]{\spacefactor3000\relax}%
\providecommand \BibitemShut  [1]{\csname bibitem#1\endcsname}%
\let\auto@bib@innerbib\@empty
\bibitem [{\citenamefont {Carvalho}\ \emph {et~al.}(2008)\citenamefont {Carvalho}, \citenamefont {Rebello}, \citenamefont {Souza}, \citenamefont {Sagrilo},\ and\ \citenamefont {Soares}}]{carvalho2008reliability}%
  \BibitemOpen
  \bibfield  {author} {\bibinfo {author} {\bibfnamefont {A.}~\bibnamefont {Carvalho}}, \bibinfo {author} {\bibfnamefont {J.}~\bibnamefont {Rebello}}, \bibinfo {author} {\bibfnamefont {M.}~\bibnamefont {Souza}}, \bibinfo {author} {\bibfnamefont {L.}~\bibnamefont {Sagrilo}},\ and\ \bibinfo {author} {\bibfnamefont {S.}~\bibnamefont {Soares}},\ }\bibfield  {title} {\bibinfo {title} {Reliability of non-destructive test techniques in the inspection of pipelines used in the oil industry},\ }\href@noop {} {\bibfield  {journal} {\bibinfo  {journal} {International journal of pressure vessels and piping}\ }\textbf {\bibinfo {volume} {85}},\ \bibinfo {pages} {745} (\bibinfo {year} {2008})}\BibitemShut {NoStop}%
\bibitem [{\citenamefont {Ma}\ \emph {et~al.}(2021)\citenamefont {Ma}, \citenamefont {Tian}, \citenamefont {Zeng}, \citenamefont {Li}, \citenamefont {Song}, \citenamefont {Wang}, \citenamefont {Gao},\ and\ \citenamefont {Zeng}}]{ma2021pipeline}%
  \BibitemOpen
  \bibfield  {author} {\bibinfo {author} {\bibfnamefont {Q.}~\bibnamefont {Ma}}, \bibinfo {author} {\bibfnamefont {G.}~\bibnamefont {Tian}}, \bibinfo {author} {\bibfnamefont {Y.}~\bibnamefont {Zeng}}, \bibinfo {author} {\bibfnamefont {R.}~\bibnamefont {Li}}, \bibinfo {author} {\bibfnamefont {H.}~\bibnamefont {Song}}, \bibinfo {author} {\bibfnamefont {Z.}~\bibnamefont {Wang}}, \bibinfo {author} {\bibfnamefont {B.}~\bibnamefont {Gao}},\ and\ \bibinfo {author} {\bibfnamefont {K.}~\bibnamefont {Zeng}},\ }\bibfield  {title} {\bibinfo {title} {Pipeline in-line inspection method, instrumentation and data management},\ }\href@noop {} {\bibfield  {journal} {\bibinfo  {journal} {Sensors}\ }\textbf {\bibinfo {volume} {21}},\ \bibinfo {pages} {3862} (\bibinfo {year} {2021})}\BibitemShut {NoStop}%
\bibitem [{\citenamefont {Coramik}\ and\ \citenamefont {Ege}(2017)}]{coramik2017discontinuity}%
  \BibitemOpen
  \bibfield  {author} {\bibinfo {author} {\bibfnamefont {M.}~\bibnamefont {Coramik}}\ and\ \bibinfo {author} {\bibfnamefont {Y.}~\bibnamefont {Ege}},\ }\bibfield  {title} {\bibinfo {title} {Discontinuity inspection in pipelines: A comparison review},\ }\href@noop {} {\bibfield  {journal} {\bibinfo  {journal} {Measurement}\ }\textbf {\bibinfo {volume} {111}},\ \bibinfo {pages} {359} (\bibinfo {year} {2017})}\BibitemShut {NoStop}%
\bibitem [{\citenamefont {Zang}\ \emph {et~al.}(2023)\citenamefont {Zang}, \citenamefont {Xu}, \citenamefont {Lu}, \citenamefont {Zhu},\ and\ \citenamefont {Zhang}}]{zang2023ultrasonic}%
  \BibitemOpen
  \bibfield  {author} {\bibinfo {author} {\bibfnamefont {X.}~\bibnamefont {Zang}}, \bibinfo {author} {\bibfnamefont {Z.-D.}\ \bibnamefont {Xu}}, \bibinfo {author} {\bibfnamefont {H.}~\bibnamefont {Lu}}, \bibinfo {author} {\bibfnamefont {C.}~\bibnamefont {Zhu}},\ and\ \bibinfo {author} {\bibfnamefont {Z.}~\bibnamefont {Zhang}},\ }\bibfield  {title} {\bibinfo {title} {Ultrasonic guided wave techniques and applications in pipeline defect detection: A review},\ }\href@noop {} {\bibfield  {journal} {\bibinfo  {journal} {International Journal of Pressure Vessels and Piping}\ ,\ \bibinfo {pages} {105033}} (\bibinfo {year} {2023})}\BibitemShut {NoStop}%
\bibitem [{\citenamefont {Ghavamian}\ \emph {et~al.}(2018)\citenamefont {Ghavamian}, \citenamefont {Mustapha}, \citenamefont {Baharudin},\ and\ \citenamefont {Yidris}}]{ghavamian2018detection}%
  \BibitemOpen
  \bibfield  {author} {\bibinfo {author} {\bibfnamefont {A.}~\bibnamefont {Ghavamian}}, \bibinfo {author} {\bibfnamefont {F.}~\bibnamefont {Mustapha}}, \bibinfo {author} {\bibfnamefont {B.~H.~T.}\ \bibnamefont {Baharudin}},\ and\ \bibinfo {author} {\bibfnamefont {N.}~\bibnamefont {Yidris}},\ }\bibfield  {title} {\bibinfo {title} {Detection, localisation and assessment of defects in pipes using guided wave techniques: A review},\ }\href@noop {} {\bibfield  {journal} {\bibinfo  {journal} {Sensors}\ }\textbf {\bibinfo {volume} {18}},\ \bibinfo {pages} {4470} (\bibinfo {year} {2018})}\BibitemShut {NoStop}%
\bibitem [{\citenamefont {Wang}\ \emph {et~al.}(2021)\citenamefont {Wang}, \citenamefont {Gao}, \citenamefont {Zhao},\ and\ \citenamefont {Wang}}]{wang2021time}%
  \BibitemOpen
  \bibfield  {author} {\bibinfo {author} {\bibfnamefont {X.}~\bibnamefont {Wang}}, \bibinfo {author} {\bibfnamefont {H.}~\bibnamefont {Gao}}, \bibinfo {author} {\bibfnamefont {K.}~\bibnamefont {Zhao}},\ and\ \bibinfo {author} {\bibfnamefont {C.}~\bibnamefont {Wang}},\ }\bibfield  {title} {\bibinfo {title} {Time-frequency characteristics of longitudinal modes in symmetric mode conversion for defect characterization in guided waves-based pipeline inspection},\ }\href@noop {} {\bibfield  {journal} {\bibinfo  {journal} {NDT \& E International}\ }\textbf {\bibinfo {volume} {122}},\ \bibinfo {pages} {102490} (\bibinfo {year} {2021})}\BibitemShut {NoStop}%
\bibitem [{\citenamefont {Lowe}\ \emph {et~al.}(2015)\citenamefont {Lowe}, \citenamefont {Sanderson}, \citenamefont {Boulgouris}, \citenamefont {Haig},\ and\ \citenamefont {Balachandran}}]{lowe2015inspection}%
  \BibitemOpen
  \bibfield  {author} {\bibinfo {author} {\bibfnamefont {P.~S.}\ \bibnamefont {Lowe}}, \bibinfo {author} {\bibfnamefont {R.~M.}\ \bibnamefont {Sanderson}}, \bibinfo {author} {\bibfnamefont {N.~V.}\ \bibnamefont {Boulgouris}}, \bibinfo {author} {\bibfnamefont {A.~G.}\ \bibnamefont {Haig}},\ and\ \bibinfo {author} {\bibfnamefont {W.}~\bibnamefont {Balachandran}},\ }\bibfield  {title} {\bibinfo {title} {Inspection of cylindrical structures using the first longitudinal guided wave mode in isolation for higher flaw sensitivity},\ }\href@noop {} {\bibfield  {journal} {\bibinfo  {journal} {IEEE Sensors Journal}\ }\textbf {\bibinfo {volume} {16}},\ \bibinfo {pages} {706} (\bibinfo {year} {2015})}\BibitemShut {NoStop}%
\bibitem [{\citenamefont {Zhang}\ \emph {et~al.}(2017)\citenamefont {Zhang}, \citenamefont {Tang}, \citenamefont {Lv},\ and\ \citenamefont {Pan}}]{zhang2017helical}%
  \BibitemOpen
  \bibfield  {author} {\bibinfo {author} {\bibfnamefont {X.}~\bibnamefont {Zhang}}, \bibinfo {author} {\bibfnamefont {Z.}~\bibnamefont {Tang}}, \bibinfo {author} {\bibfnamefont {F.}~\bibnamefont {Lv}},\ and\ \bibinfo {author} {\bibfnamefont {X.}~\bibnamefont {Pan}},\ }\bibfield  {title} {\bibinfo {title} {Helical comb magnetostrictive patch transducers for inspecting spiral welded pipes using flexural guided waves},\ }\href@noop {} {\bibfield  {journal} {\bibinfo  {journal} {Ultrasonics}\ }\textbf {\bibinfo {volume} {74}},\ \bibinfo {pages} {1} (\bibinfo {year} {2017})}\BibitemShut {NoStop}%
\bibitem [{\citenamefont {Kim}\ and\ \citenamefont {Kwon}(2015)}]{kim2015review}%
  \BibitemOpen
  \bibfield  {author} {\bibinfo {author} {\bibfnamefont {Y.~Y.}\ \bibnamefont {Kim}}\ and\ \bibinfo {author} {\bibfnamefont {Y.~E.}\ \bibnamefont {Kwon}},\ }\bibfield  {title} {\bibinfo {title} {Review of magnetostrictive patch transducers and applications in ultrasonic nondestructive testing of waveguides},\ }\href@noop {} {\bibfield  {journal} {\bibinfo  {journal} {Ultrasonics}\ }\textbf {\bibinfo {volume} {62}},\ \bibinfo {pages} {3} (\bibinfo {year} {2015})}\BibitemShut {NoStop}%
\bibitem [{\citenamefont {Niu}\ \emph {et~al.}(2019)\citenamefont {Niu}, \citenamefont {Duan}, \citenamefont {Chen},\ and\ \citenamefont {Marques}}]{niu2019excitation}%
  \BibitemOpen
  \bibfield  {author} {\bibinfo {author} {\bibfnamefont {X.}~\bibnamefont {Niu}}, \bibinfo {author} {\bibfnamefont {W.}~\bibnamefont {Duan}}, \bibinfo {author} {\bibfnamefont {H.-P.}\ \bibnamefont {Chen}},\ and\ \bibinfo {author} {\bibfnamefont {H.}~\bibnamefont {Marques}},\ }\bibfield  {title} {\bibinfo {title} {Excitation and propagation of torsional t (0, 1) mode for guided wave testing of pipeline integrity},\ }\href@noop {} {\bibfield  {journal} {\bibinfo  {journal} {Measurement}\ }\textbf {\bibinfo {volume} {131}},\ \bibinfo {pages} {341} (\bibinfo {year} {2019})}\BibitemShut {NoStop}%
\bibitem [{\citenamefont {Kim}\ \emph {et~al.}(2013)\citenamefont {Kim}, \citenamefont {Lee},\ and\ \citenamefont {Kim}}]{kim2013circumferential}%
  \BibitemOpen
  \bibfield  {author} {\bibinfo {author} {\bibfnamefont {H.~W.}\ \bibnamefont {Kim}}, \bibinfo {author} {\bibfnamefont {J.~K.}\ \bibnamefont {Lee}},\ and\ \bibinfo {author} {\bibfnamefont {Y.~Y.}\ \bibnamefont {Kim}},\ }\bibfield  {title} {\bibinfo {title} {Circumferential phased array of shear-horizontal wave magnetostrictive patch transducers for pipe inspection},\ }\href@noop {} {\bibfield  {journal} {\bibinfo  {journal} {Ultrasonics}\ }\textbf {\bibinfo {volume} {53}},\ \bibinfo {pages} {423} (\bibinfo {year} {2013})}\BibitemShut {NoStop}%
\bibitem [{\citenamefont {Miao}\ \emph {et~al.}(2017)\citenamefont {Miao}, \citenamefont {Huan}, \citenamefont {Wang},\ and\ \citenamefont {Li}}]{miao2017excitation}%
  \BibitemOpen
  \bibfield  {author} {\bibinfo {author} {\bibfnamefont {H.}~\bibnamefont {Miao}}, \bibinfo {author} {\bibfnamefont {Q.}~\bibnamefont {Huan}}, \bibinfo {author} {\bibfnamefont {Q.}~\bibnamefont {Wang}},\ and\ \bibinfo {author} {\bibfnamefont {F.}~\bibnamefont {Li}},\ }\bibfield  {title} {\bibinfo {title} {Excitation and reception of single torsional wave t (0, 1) mode in pipes using face-shear d24 piezoelectric ring array},\ }\href@noop {} {\bibfield  {journal} {\bibinfo  {journal} {Smart Materials and Structures}\ }\textbf {\bibinfo {volume} {26}},\ \bibinfo {pages} {025021} (\bibinfo {year} {2017})}\BibitemShut {NoStop}%
\bibitem [{\citenamefont {Zhou}\ \emph {et~al.}(2016)\citenamefont {Zhou}, \citenamefont {Yuan},\ and\ \citenamefont {Shi}}]{zhou2016guided}%
  \BibitemOpen
  \bibfield  {author} {\bibinfo {author} {\bibfnamefont {W.}~\bibnamefont {Zhou}}, \bibinfo {author} {\bibfnamefont {F.-G.}\ \bibnamefont {Yuan}},\ and\ \bibinfo {author} {\bibfnamefont {T.}~\bibnamefont {Shi}},\ }\bibfield  {title} {\bibinfo {title} {Guided torsional wave generation of a linear in-plane shear piezoelectric array in metallic pipes},\ }\href@noop {} {\bibfield  {journal} {\bibinfo  {journal} {Ultrasonics}\ }\textbf {\bibinfo {volume} {65}},\ \bibinfo {pages} {69} (\bibinfo {year} {2016})}\BibitemShut {NoStop}%
\bibitem [{\citenamefont {Heinlein}\ \emph {et~al.}(2018)\citenamefont {Heinlein}, \citenamefont {Cawley},\ and\ \citenamefont {Vogt}}]{heinlein2018reflection}%
  \BibitemOpen
  \bibfield  {author} {\bibinfo {author} {\bibfnamefont {S.}~\bibnamefont {Heinlein}}, \bibinfo {author} {\bibfnamefont {P.}~\bibnamefont {Cawley}},\ and\ \bibinfo {author} {\bibfnamefont {T.}~\bibnamefont {Vogt}},\ }\bibfield  {title} {\bibinfo {title} {{Reflection of torsional T(0,1) guided waves from defects in pipe bends}},\ }\href@noop {} {\bibfield  {journal} {\bibinfo  {journal} {NDT \& E International}\ }\textbf {\bibinfo {volume} {93}},\ \bibinfo {pages} {57} (\bibinfo {year} {2018})}\BibitemShut {NoStop}%
\bibitem [{\citenamefont {L{\o}vstad}\ and\ \citenamefont {Cawley}(2011)}]{lovstad2011reflection}%
  \BibitemOpen
  \bibfield  {author} {\bibinfo {author} {\bibfnamefont {A.}~\bibnamefont {L{\o}vstad}}\ and\ \bibinfo {author} {\bibfnamefont {P.}~\bibnamefont {Cawley}},\ }\bibfield  {title} {\bibinfo {title} {The reflection of the fundamental torsional guided wave from multiple circular holes in pipes},\ }\href@noop {} {\bibfield  {journal} {\bibinfo  {journal} {NDT \& E International}\ }\textbf {\bibinfo {volume} {44}},\ \bibinfo {pages} {553} (\bibinfo {year} {2011})}\BibitemShut {NoStop}%
\bibitem [{\citenamefont {Fletcher}\ \emph {et~al.}(2012)\citenamefont {Fletcher}, \citenamefont {Lowe}, \citenamefont {Ratassepp},\ and\ \citenamefont {Brett}}]{fletcher2012detection}%
  \BibitemOpen
  \bibfield  {author} {\bibinfo {author} {\bibfnamefont {S.}~\bibnamefont {Fletcher}}, \bibinfo {author} {\bibfnamefont {M.~J.}\ \bibnamefont {Lowe}}, \bibinfo {author} {\bibfnamefont {M.}~\bibnamefont {Ratassepp}},\ and\ \bibinfo {author} {\bibfnamefont {C.}~\bibnamefont {Brett}},\ }\bibfield  {title} {\bibinfo {title} {Detection of axial cracks in pipes using focused guided waves},\ }\href@noop {} {\bibfield  {journal} {\bibinfo  {journal} {Journal of Nondestructive Evaluation}\ }\textbf {\bibinfo {volume} {31}},\ \bibinfo {pages} {56} (\bibinfo {year} {2012})}\BibitemShut {NoStop}%
\bibitem [{\citenamefont {Liu}\ \emph {et~al.}(2006)\citenamefont {Liu}, \citenamefont {He}, \citenamefont {Wu}, \citenamefont {Wang},\ and\ \citenamefont {Yang}}]{liu2006circumferential}%
  \BibitemOpen
  \bibfield  {author} {\bibinfo {author} {\bibfnamefont {Z.}~\bibnamefont {Liu}}, \bibinfo {author} {\bibfnamefont {C.}~\bibnamefont {He}}, \bibinfo {author} {\bibfnamefont {B.}~\bibnamefont {Wu}}, \bibinfo {author} {\bibfnamefont {X.}~\bibnamefont {Wang}},\ and\ \bibinfo {author} {\bibfnamefont {S.}~\bibnamefont {Yang}},\ }\bibfield  {title} {\bibinfo {title} {Circumferential and longitudinal defect detection using t (0, 1) mode excited by thickness shear mode piezoelectric elements},\ }\href@noop {} {\bibfield  {journal} {\bibinfo  {journal} {Ultrasonics}\ }\textbf {\bibinfo {volume} {44}},\ \bibinfo {pages} {e1135} (\bibinfo {year} {2006})}\BibitemShut {NoStop}%
\bibitem [{\citenamefont {Yeung}\ and\ \citenamefont {Ng}(2019)}]{yeung2019time}%
  \BibitemOpen
  \bibfield  {author} {\bibinfo {author} {\bibfnamefont {C.}~\bibnamefont {Yeung}}\ and\ \bibinfo {author} {\bibfnamefont {C.~T.}\ \bibnamefont {Ng}},\ }\bibfield  {title} {\bibinfo {title} {Time-domain spectral finite element method for analysis of torsional guided waves scattering and mode conversion by cracks in pipes},\ }\href@noop {} {\bibfield  {journal} {\bibinfo  {journal} {Mechanical Systems and Signal Processing}\ }\textbf {\bibinfo {volume} {128}},\ \bibinfo {pages} {305} (\bibinfo {year} {2019})}\BibitemShut {NoStop}%
\bibitem [{\citenamefont {Cawley}(2021)}]{cawley2021development}%
  \BibitemOpen
  \bibfield  {author} {\bibinfo {author} {\bibfnamefont {P.}~\bibnamefont {Cawley}},\ }\bibfield  {title} {\bibinfo {title} {A development strategy for structural health monitoring applications},\ }\href@noop {} {\bibfield  {journal} {\bibinfo  {journal} {Journal of Nondestructive Evaluation, Diagnostics and Prognostics of Engineering Systems}\ }\textbf {\bibinfo {volume} {4}},\ \bibinfo {pages} {041012} (\bibinfo {year} {2021})}\BibitemShut {NoStop}%
\bibitem [{\citenamefont {Sawant}\ \emph {et~al.}(2022)\citenamefont {Sawant}, \citenamefont {Patil}, \citenamefont {Thalapil}, \citenamefont {Banerjee},\ and\ \citenamefont {Tallur}}]{sawant2022temperature}%
  \BibitemOpen
  \bibfield  {author} {\bibinfo {author} {\bibfnamefont {S.}~\bibnamefont {Sawant}}, \bibinfo {author} {\bibfnamefont {S.}~\bibnamefont {Patil}}, \bibinfo {author} {\bibfnamefont {J.}~\bibnamefont {Thalapil}}, \bibinfo {author} {\bibfnamefont {S.}~\bibnamefont {Banerjee}},\ and\ \bibinfo {author} {\bibfnamefont {S.}~\bibnamefont {Tallur}},\ }\bibfield  {title} {\bibinfo {title} {{Temperature variation compensated damage classification and localisation in ultrasonic guided wave SHM using self-learnt features and Gaussian mixture models}},\ }\href@noop {} {\bibfield  {journal} {\bibinfo  {journal} {Smart Materials and Structures}\ }\textbf {\bibinfo {volume} {31}},\ \bibinfo {pages} {055008} (\bibinfo {year} {2022})}\BibitemShut {NoStop}%
\bibitem [{\citenamefont {Ramana}\ \emph {et~al.}(2023)\citenamefont {Ramana}, \citenamefont {Patil}, \citenamefont {Kashyap}, \citenamefont {Tallur},\ and\ \citenamefont {Banerjee}}]{ramana2023effect}%
  \BibitemOpen
  \bibfield  {author} {\bibinfo {author} {\bibfnamefont {R.~B.}\ \bibnamefont {Ramana}}, \bibinfo {author} {\bibfnamefont {S.}~\bibnamefont {Patil}}, \bibinfo {author} {\bibfnamefont {P.}~\bibnamefont {Kashyap}}, \bibinfo {author} {\bibfnamefont {S.}~\bibnamefont {Tallur}},\ and\ \bibinfo {author} {\bibfnamefont {S.}~\bibnamefont {Banerjee}},\ }\bibfield  {title} {\bibinfo {title} {{The effect of temperature on guided wave signal characteristics in presence of disbond and delamination for health monitoring of a honeycomb composite sandwich structure with built-in PZT network}},\ }\href@noop {} {\bibfield  {journal} {\bibinfo  {journal} {Smart Materials and Structures}\ }\textbf {\bibinfo {volume} {32}},\ \bibinfo {pages} {095003} (\bibinfo {year} {2023})}\BibitemShut {NoStop}%
\bibitem [{\citenamefont {Kashyap}\ \emph {et~al.}(2023)\citenamefont {Kashyap}, \citenamefont {Shivgan}, \citenamefont {Patil}, \citenamefont {Mahajan}, \citenamefont {Banerjee},\ and\ \citenamefont {Tallur}}]{kashyap2023tinyml}%
  \BibitemOpen
  \bibfield  {author} {\bibinfo {author} {\bibfnamefont {P.}~\bibnamefont {Kashyap}}, \bibinfo {author} {\bibfnamefont {K.}~\bibnamefont {Shivgan}}, \bibinfo {author} {\bibfnamefont {S.}~\bibnamefont {Patil}}, \bibinfo {author} {\bibfnamefont {S.}~\bibnamefont {Mahajan}}, \bibinfo {author} {\bibfnamefont {S.}~\bibnamefont {Banerjee}},\ and\ \bibinfo {author} {\bibfnamefont {S.}~\bibnamefont {Tallur}},\ }\bibfield  {title} {\bibinfo {title} {{TinyML-enabled unsupervised ultrasonic guided wave SHM under varying thermal conditions}},\ }in\ \href@noop {} {\emph {\bibinfo {booktitle} {2023 Joint Conference of the European Frequency and Time Forum and IEEE International Frequency Control Symposium (EFTF/IFCS)}}}\ (\bibinfo {organization} {IEEE},\ \bibinfo {year} {2023})\ pp.\ \bibinfo {pages} {1--3}\BibitemShut {NoStop}%
\bibitem [{\citenamefont {Kashyap}\ \emph {et~al.}(2024)\citenamefont {Kashyap}, \citenamefont {Shivgan}, \citenamefont {Patil}, \citenamefont {Raja}, \citenamefont {Mahajan}, \citenamefont {Banerjee},\ and\ \citenamefont {Tallur}}]{kashyap2024unsupervised}%
  \BibitemOpen
  \bibfield  {author} {\bibinfo {author} {\bibfnamefont {P.}~\bibnamefont {Kashyap}}, \bibinfo {author} {\bibfnamefont {K.}~\bibnamefont {Shivgan}}, \bibinfo {author} {\bibfnamefont {S.}~\bibnamefont {Patil}}, \bibinfo {author} {\bibfnamefont {B.~R.}\ \bibnamefont {Raja}}, \bibinfo {author} {\bibfnamefont {S.}~\bibnamefont {Mahajan}}, \bibinfo {author} {\bibfnamefont {S.}~\bibnamefont {Banerjee}},\ and\ \bibinfo {author} {\bibfnamefont {S.}~\bibnamefont {Tallur}},\ }\bibfield  {title} {\bibinfo {title} {Unsupervised deep learning framework for temperature-compensated damage assessment using ultrasonic guided waves on edge device},\ }\href@noop {} {\bibfield  {journal} {\bibinfo  {journal} {Scientific Reports}\ }\textbf {\bibinfo {volume} {14}},\ \bibinfo {pages} {3751} (\bibinfo {year} {2024})}\BibitemShut {NoStop}%
\bibitem [{\citenamefont {Satyanarayan}\ \emph {et~al.}(2007)\citenamefont {Satyanarayan}, \citenamefont {Sridhar}, \citenamefont {Krishnamurthy},\ and\ \citenamefont {Balasubramaniam}}]{satyanarayan2007simulation}%
  \BibitemOpen
  \bibfield  {author} {\bibinfo {author} {\bibfnamefont {L.}~\bibnamefont {Satyanarayan}}, \bibinfo {author} {\bibfnamefont {C.}~\bibnamefont {Sridhar}}, \bibinfo {author} {\bibfnamefont {C.}~\bibnamefont {Krishnamurthy}},\ and\ \bibinfo {author} {\bibfnamefont {K.}~\bibnamefont {Balasubramaniam}},\ }\bibfield  {title} {\bibinfo {title} {Simulation of ultrasonic phased array technique for imaging and sizing of defects using longitudinal waves},\ }\href@noop {} {\bibfield  {journal} {\bibinfo  {journal} {International Journal of Pressure Vessels and Piping}\ }\textbf {\bibinfo {volume} {84}},\ \bibinfo {pages} {716} (\bibinfo {year} {2007})}\BibitemShut {NoStop}%
\bibitem [{\citenamefont {Zhou}\ \emph {et~al.}(2018)\citenamefont {Zhou}, \citenamefont {Qian}, \citenamefont {Birnie},\ and\ \citenamefont {Zhao}}]{zhou2018reference}%
  \BibitemOpen
  \bibfield  {author} {\bibinfo {author} {\bibfnamefont {Y.-L.}\ \bibnamefont {Zhou}}, \bibinfo {author} {\bibfnamefont {X.}~\bibnamefont {Qian}}, \bibinfo {author} {\bibfnamefont {A.}~\bibnamefont {Birnie}},\ and\ \bibinfo {author} {\bibfnamefont {X.-L.}\ \bibnamefont {Zhao}},\ }\bibfield  {title} {\bibinfo {title} {A reference free ultrasonic phased array to identify surface cracks in welded steel pipes based on transmissibility},\ }\href@noop {} {\bibfield  {journal} {\bibinfo  {journal} {International Journal of Pressure Vessels and Piping}\ }\textbf {\bibinfo {volume} {168}},\ \bibinfo {pages} {66} (\bibinfo {year} {2018})}\BibitemShut {NoStop}%
\bibitem [{\citenamefont {Davies}\ and\ \citenamefont {Cawley}(2009)}]{davies2009application}%
  \BibitemOpen
  \bibfield  {author} {\bibinfo {author} {\bibfnamefont {J.}~\bibnamefont {Davies}}\ and\ \bibinfo {author} {\bibfnamefont {P.}~\bibnamefont {Cawley}},\ }\bibfield  {title} {\bibinfo {title} {The application of synthetic focusing for imaging crack-like defects in pipelines using guided waves},\ }\href@noop {} {\bibfield  {journal} {\bibinfo  {journal} {IEEE transactions on ultrasonics, ferroelectrics, and frequency control}\ }\textbf {\bibinfo {volume} {56}},\ \bibinfo {pages} {759} (\bibinfo {year} {2009})}\BibitemShut {NoStop}%
\bibitem [{\citenamefont {Davies}\ \emph {et~al.}(2006)\citenamefont {Davies}, \citenamefont {Simonetti}, \citenamefont {Lowe},\ and\ \citenamefont {Cawley}}]{davies2006review}%
  \BibitemOpen
  \bibfield  {author} {\bibinfo {author} {\bibfnamefont {J.}~\bibnamefont {Davies}}, \bibinfo {author} {\bibfnamefont {F.}~\bibnamefont {Simonetti}}, \bibinfo {author} {\bibfnamefont {M.}~\bibnamefont {Lowe}},\ and\ \bibinfo {author} {\bibfnamefont {P.}~\bibnamefont {Cawley}},\ }\bibfield  {title} {\bibinfo {title} {Review of synthetically focused guided wave imaging techniques with application to defect sizing},\ }in\ \href@noop {} {\emph {\bibinfo {booktitle} {AIP conference proceedings}}},\ Vol.\ \bibinfo {volume} {820}\ (\bibinfo {organization} {American Institute of Physics},\ \bibinfo {year} {2006})\ pp.\ \bibinfo {pages} {142--149}\BibitemShut {NoStop}%
\bibitem [{\citenamefont {Patil}\ \emph {et~al.}(2023)\citenamefont {Patil}, \citenamefont {Mahajan}, \citenamefont {Banerjee},\ and\ \citenamefont {Tallur}}]{patil2023embedded}%
  \BibitemOpen
  \bibfield  {author} {\bibinfo {author} {\bibfnamefont {S.}~\bibnamefont {Patil}}, \bibinfo {author} {\bibfnamefont {S.}~\bibnamefont {Mahajan}}, \bibinfo {author} {\bibfnamefont {S.}~\bibnamefont {Banerjee}},\ and\ \bibinfo {author} {\bibfnamefont {S.}~\bibnamefont {Tallur}},\ }\bibfield  {title} {\bibinfo {title} {An embedded shm system for monitoring of pipelines using torsional guided wave ultrasonics},\ }in\ \href@noop {} {\emph {\bibinfo {booktitle} {2023 Joint Conference of the European Frequency and Time Forum and IEEE International Frequency Control Symposium (EFTF/IFCS)}}}\ (\bibinfo {organization} {IEEE},\ \bibinfo {year} {2023})\ pp.\ \bibinfo {pages} {1--3}\BibitemShut {NoStop}%
\end{thebibliography}%

\end{document}